\pdfoutput=1

\documentclass[11pt]{article}

\usepackage{ACL2023}

\usepackage{times}
\usepackage{latexsym}

\usepackage[T1]{fontenc}

\usepackage[utf8]{inputenc}

\usepackage{microtype}

\usepackage{inconsolata}

\usepackage{amsmath,amsfonts,bm}

\def\eqref#1{equation~\ref{#1}}

\def\1{\bm{1}}

\def\rve{{\mathbf{e}}}

\def\rvp{{\mathbf{p}}}

\def\rvs{{\mathbf{s}}}

\def\rvx{{\mathbf{x}}}
\def\rvy{{\mathbf{y}}}
\def\rvz{{\mathbf{z}}}

\def\rmE{{\mathbf{E}}}

\def\rmX{{\mathbf{X}}}

\def\rmZ{{\mathbf{Z}}}

\DeclareMathAlphabet{\mathsfit}{\encodingdefault}{\sfdefault}{m}{sl}
\SetMathAlphabet{\mathsfit}{bold}{\encodingdefault}{\sfdefault}{bx}{n}

\DeclareMathOperator*{\argmax}{arg\,max}

\usepackage{hyperref}
\usepackage{url}
\usepackage{amsmath,graphicx}
\usepackage{multirow}
\usepackage{makecell}
\usepackage{booktabs}
\usepackage{subcaption}
\usepackage{amssymb}
\usepackage{amsmath}
\usepackage{graphicx}
\usepackage{enumitem}
\usepackage{threeparttable}
\usepackage{diagbox}
\usepackage{setspace}
\usepackage{array,caption}
\usepackage[labelfont=bf]{caption}

\newcommand{\change}[1]{{{#1}}}

\newcommand{{\rpc}}{{RepCodec}}
\usepackage[capitalize,noabbrev]{cleveref}

\usepackage{color}

\title{RepCodec: A Speech Representation Codec for Speech Tokenization}

\author{Zhichao Huang\thanks{\thickspace \thickspace Contribute equally}\quad\ Chutong Meng\footnotemark[1] \quad Tom Ko
\thanks{\thickspace \thickspace Corresponding author} 
\\
ByteDance\\
\texttt{\{zhichao.huang, tom.ko\}@bytedance.com, mengct00@gmail.com}
}

\begin{document}
\maketitle
\begin{abstract}

With recent rapid growth of large language models (LLMs), discrete speech tokenization has played an important role for injecting speech into LLMs. However, this discretization gives rise to a loss of information, consequently impairing overall performance. To improve the performance of these discrete speech tokens, we present {\rpc}, a novel speech representation codec for semantic speech tokenization. In contrast to audio codecs which reconstruct the raw audio, {\rpc} learns a vector quantization codebook through reconstructing speech representations from speech encoders like HuBERT or data2vec. 
Together, the speech encoder, the codec encoder and the vector quantization codebook form a pipeline for converting speech waveforms into semantic tokens. 
The extensive experiments illustrate that {\rpc}, by virtue of its enhanced information retention capacity, significantly outperforms the widely used k-means clustering approach in both speech understanding and generation. Furthermore, this superiority extends across various speech encoders and languages, affirming the robustness of {\rpc}.
We believe our method can facilitate large language modeling research on speech processing. Our code and models are released at \url{https://github.com/mct10/RepCodec}.

\end{abstract}
\section{Introduction}
\label{sec:intro}

The significant achievements of large language models (LLMs) within the field of natural language processing have attracted considerable attention, as evidenced by notable works such as \citet{openai2023gpt4, brown2020languagegpt3, radford2019languagegpt2, wei2021finetunedflan, chowdhery2022palm}.
Bridging the realms of continuous speech and token-based language models necessitates a key technique known as speech tokenization, which discretizes an audio signal into a finite set of tokens.
By converting speech into discrete tokens, language models can predict the future semantic content and generate realistic speech with long-term consistency \citep{nguyen2022discrete}.
As a result, a growing body of research has begun to incorporate speech tokenization into the realm of LLM. Noteworthy examples include AudioLM \citep{borsos2023audiolm}, AudioPaLM \citep{rubenstein2023audiopalm}, Vall-E \citep{wang2023neural}, PolyVoice \citep{dong2023polyvoice}, and SpeechGPT \citep{zhang2023speechgpt}.

Discrete speech tokens can be divided into two categories: semantic tokens and acoustic tokens. 
Acoustic tokens are produced by audio codecs \change{\citep{soundstream, defossez2022high, zhang2023speechtokenizer}}, which aim to reconstruct the original audio so that it can be perceptually identical to listeners. 
However, attempting to preserve all information of the audios leads to high bitrates of acoustic tokens. 
The process not only imposes significant computational demands on the LLMs, but sometimes makes training infeasible with such lengthy sequences.
For example, converting a 30-second audio segment into acoustic tokens needs 18,000 tokens. 
Therefore, current language modeling approaches often require substantial architectural adjustments to accommodate such long sequences \citep{borsos2023audiolm,wang2023neural}.

Semantic tokens, on the other hand, aim at preserving only the semantic information of the audio, which allows much lower bitrates. 
If the task relies only on the content of the speech (\textit{e.g.} speech recognition / translation), using semantic tokens should be a better choice.
At present, k-means clustering on speech representations \citep{hsu2021hubert} is the most prevalent technique of extracting semantic tokens.
However, this method has two drawbacks. 
Firstly, it suffers from a loss of semantic information compared to the original speech representations \citep{lee-etal-2022-textless, borsos2023audiolm}. 
Secondly, not all sets of speech representations are suitable for clustering.
\citet{rubenstein2023audiopalm} reported that the choice of speech encoder significantly affects the downstream task performance of speech tokenization.
Thus, we are motivated to improve the method of extracting semantic tokens by addressing the above problems.

\begin{figure*}[t]
\centering
\centerline{\includegraphics[width=\linewidth]{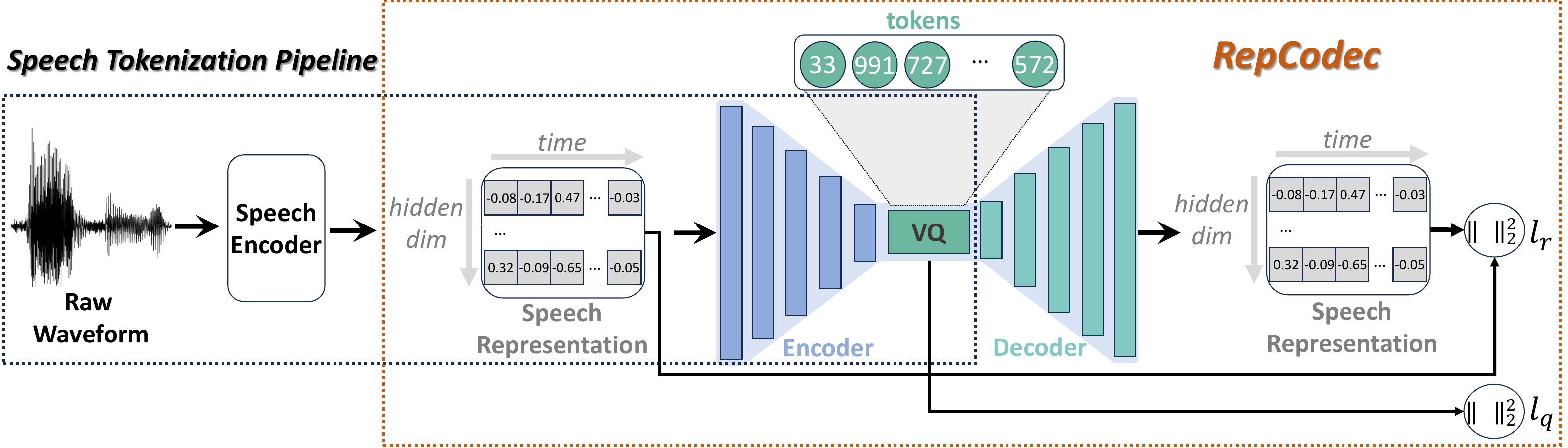}}
\caption{{\rpc} model architecture. \change{Our network uses single residual units without dimension reduction.}}
\label{fig:repcodec}
\end{figure*}

In this paper, we propose {\rpc}, a general tokenization approach for representations that can be applied to speech to extract its semantic tokens.
{\rpc} leverages an end-to-end neural codec to preserve more information of the speech representations. {\rpc} is composed of an encoder, a vector quantizer (VQ) and a decoder, and it is trained to reconstruct the input speech representations as much as possible.
The speech encoder, codec encoder and the VQ codebook together constitute the speech tokenization pipeline, which can produce high-quality semantic tokens for downstream tasks with a low bitrate.
We evaluate the {\rpc} tokens in downstream tasks of speech understanding and generation. Specifically, we use a decoder-only ASR modelling task to evaluate the speech understanding of {\rpc} tokens, and a unit-to-speech generation task to measure the quality of {\rpc} tokens for speech generation. Our comprehensive experiments demonstrate that {\rpc} significantly outperforms the dominant k-means clustering approach.
It is worth noting that supervised approach such as ASR and phoneme recognition can also be considered as forms of speech tokenization that convert speech into word tokens or phoneme tokens. However, it is essential to highlight the large amount of parallel data required for supervised training only exists for high-resource languages. {\rpc}, on the contrary, is an unsupervised speech tokenization method that can be applied to any languages. 
The contribution of our work can be summarized as follows:

1. We propose a novel framework, {\rpc}, which applies compression techniques on representations to enhance the preservation of information within representations. 

2. The experiments show that, by applying {\rpc} to speech, semantic tokens exhibit an improved capacity for retaining information, and they surpass the prevailing k-means clustering approach in both speech understanding (4.5\% v.s 2.8\% word error rate (WER)) and generation (7.6\% v.s 4.7\% WER). In addition, further experiments demonstrate that {\rpc} is a robust method that can be applied to various speech encoders and languages.

3. Our further analysis underscores that the quality of semantic tokens primarily relies on the information loss instead of their similarity to the phonemes. This finding serves as motivation for future advancements in semantic token refinement. 





\section{Related Work}

Several lines of works are related to {\rpc}, including self-supervised speech representation learning, speech tokenization and vector quantization.

\noindent\textbf{Self-supervised Speech Representation Learning.} This line of research has recently gained huge success in the area of speech processing. Prevalent methods usually 
requires the model to predict the content of unseen regions \citep{hsu2021hubert, baevski2022data2vec, chung2021w2v, chung2020generative} or to contrast the target unseen frame with randomly sampled ones \citep{baevski2020wav2vec2, conneau2020unsupervisedXLSR}. Specifically, HuBERT \citep{hsu2021hubert} is a pioneering work to employ k-means for speech tokenization.
The generated tokens serve as the training targets for the speech encoders. And they find that these tokens exhibit a strong correlation with the phonemes. 
Later on, \citet{lee-etal-2022-textless, meng2023cobert} show that these semantic tokens can be used directly to perform downstream tasks like ASR or voice conversion.  
In addition to the self-supervised approaches that only use unlabeled speech data, many works train the speech encoder in an end-to-end manner \citep{gulati2020conformer, radford2023whisper} with labeled speech data. Among them, Whisper \citep{radford2023whisper} is currently the largest open-source model for speech recognition, which achieves the lowest overall WER across various domains.

\noindent\textbf{Speech Tokenization.} 
Discrete speech tokens can be divided into two categories: \textit{Semantic tokens} \citep{lee-etal-2022-textless, hsu2021hubert, chung2021w2v} and \textit{Acoustic tokens} \citep{soundstream, defossez2022high, wu2023audiodec}. Semantic tokens maintain the linguistic information of the speech and have high correlation with phonemes \citep{hsu2021hubert}. They are commonly generated by applying k-means clustering to pretrained speech encoders like HuBERT \citep{hsu2021hubert} or data2vec \citep{baevski2022data2vec}. Semantic tokens are widely used for downstream tasks. For example,  \citet{lee-etal-2022-textless, polyak2021speech, directs2st} use them to train a unit-vocoder to generate raw speech. AudioLM \citep{borsos2023audiolm} inputs the semantic tokens of w2v-BERT to represent semantic information of the audio. \citet{zhang2023speechgpt} jointly trains a language model with semantic tokens to inject speech recognition ability to GPT-like models. \citet{dong2023polyvoice} also incorporates semantic tokens for speech to speech translation. However, the discretization step of k-means discards plenty of information of the speech, resulting in degraded performance in downstream tasks \citep{lee-etal-2022-textless, borsos2023audiolm}. 

In contrast to semantic tokens, acoustic tokens aim to preserve all the information of the audio. Soundstream \citep{soundstream} and EnCodec \citep{defossez2022high} use a neural audio codec with Residual Vector Quantizers (RVQ) to learn acoustic tokens that can be directly reconstructed into raw audios. As these acoustic tokens contain acoustic information of the audio, they can be used to perform more complicated tasks than semantic tokens. For example, VALL-E \citep{wang2023neural}, uses them for zero-shot text to speech (TTS), and AudioLM \citep{borsos2023audiolm} employs these tokens not only to produce realistic speech but also music. However, as acoustic tokens need to preserve a lot of information unrelated to semantics, their bitrates surpass the capacity of conventional language models.Consequently, handling these tokens requires specialized techniques \citep{van2017vqvae, borsos2023audiolm}, making their practical utility challenging.

\noindent\textbf{Vector Quantization.} Learning vector quantization is important for efficient coding of information. VQ-VAE \citep{van2017vqvae} introduces vector quantization into VAE \citep{kingma2014vae} to reconstruct images by learning discrete codebooks. They propose a straight-through gradient method to allow gradient back-propagation through a non-differentiable quantization operation so that optimization of the network is feasible. 
Furthermore, recent advancements include softmax quantization \citep{kankanahalli2018end}, exponential moving average
(EMA) \citep{garbacea2019low}, and Gumble-softmax \citep{yang2022audio} also work well for optimization of the VQ module. 
In addition to multiple VQ codebooks in \citet{van2017vqvae}, Soundstream \citep{soundstream} introduces a new RVQ method that is able to compress the raw audio with different bitrates. 


\begin{figure*}[t]
\centering
\centerline{\includegraphics[width=\linewidth]{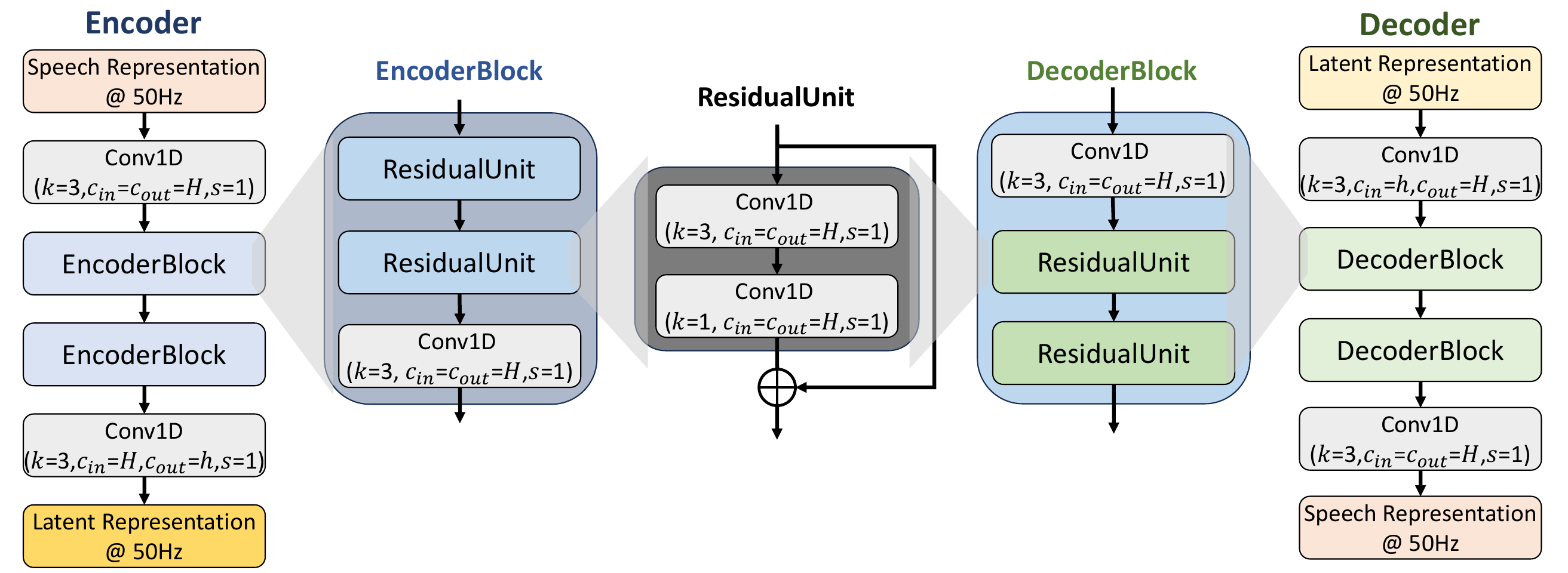}}
\caption{
Encoder and decoder architecture of {\rpc}.
$k$,  $s$,  $c_{in}$, $c_{out}$ and  denote kernel size, stride, input and output channels, $h$ denotes the number of clusters, and $H$ denotes the hidden dimension of input representations.
}
\label{fig:repcodec_arch}
\end{figure*}

\section{Method}

Despite the wide applications of semantic tokens in speech modeling \citep{borsos2023audiolm, lee-etal-2022-textless}, the discretization of the representations results in severe information loss. Consequently, downstream tasks, such as ASR or speech translation, suffer from a significant downgrade in performance. In AudioLM \citep{borsos2023audiolm}, the WER is dramatically increased from 2.5\% to 6.0\% by using the discrete tokens of k-means from w2v-BERT XL \citep{chung2021w2v}. The unit-vocoder in mHuBERT \citep{lee-etal-2022-textless} also relatively increases the WER of the generated audio by about 70\%. These results all demonstrate that the severe information loss actually prevents the discretization of speech from obtaining SOTA performance. 

The information loss motivates us to preserve more information during the discretization of the representations.
To this end, we propose a novel method, {\rpc}, to perform more efficient compression on the representations so that the semantic tokens can preserve more information and achieve better performance in downstream tasks.

\subsection{Architecture of {\rpc}}

In order to achieve better compression of the representation, {\rpc} uses a parametric network, which consists of 3 components (Figure \ref{fig:repcodec}): a codec encoder, a VQ module, and a codec decoder.
The codec encoder takes as input the speech representations $\rmX = [\rvx_1, \cdots, \rvx_T] \in \mathbb{R}^{H \times T}$ and produces latent representations $\rmZ = [\rvz_1, \cdots, \rvz_T] \in \mathbb{R}^{H \times T}$. Here, $H$ is the dimension of the speech representation and $T$ is the length of the sequence. 
$\rmZ$ is then passed through the VQ module to be quantized into a sequence of discrete tokens $\rvs = s_1\cdots s_T$ with codebook $\rmE = [\rve_1, \cdots, \rve_K]$, where $K$ is a predetermined number of clusters.
The codec decoder utilizes these tokens $\rmE$ to reconstruct the original speech representations.

\noindent \textbf{Encoder and Decoder.} The architecture of the encoder and decoder follows \citet{soundstream, wu2023audiodec}, which achieves great success in compressing audio signals. As shown in \Cref{fig:repcodec_arch}, the encoder consists of several 1D convolution layers with convolution applied to the time dimension of the input representation $\rmX$. The encoder block contains residual path to allow better optimization of the network \citep{he2016deep}. The decoder has a similar design, which is also composed of several 1D convolution layers and residual paths. In our paper, we do not downsample or upsample in both the encoder and the decoder, and keep frequency of the representation the same as the input.

\noindent\textbf{Vector Quantizer.} Vector Quantizer aims to compress the latent representations $\rmZ$ to a series of discrete tokens $\rvs$. It projects the latent $\rvz$ to its closest codebook $\rve_k$ and outputs $\rve_k$ to the decoder. We adopt two kinds of quantizer, regular VQ  \citep{van2017vqvae} and RVQ \citep{soundstream}.
RVQ is a $M$-layer quantizer where each layer quantizes the residual of the previous layer, and it is effective for compressing the audio signals. When $M=1$, RVQ is equivalent to VQ.

%




\subsection{Training Objective}
\label{sec:train_obj}
Our training objectives consist of a reconstruction loss on $\rmX$, which aims to preserve as much input information as possible for downstream tasks, and a quantization loss to effectively train the VQ. 

\noindent \textbf{Reconstruction loss $l_r$.} 
We minimize the squared $\ell_2$ distance between the input representations $\rmX$ and the output representations $\hat{\rmX}$. Formally,
\begin{equation}
    l_r = \frac{1}{HT} \|\rmX - \hat{\rmX}\|_F^2
\end{equation}
where $H$ is the hidden dimension of the representation and $\|\cdot\|_F$ denotes the Frobenius norm.

\noindent \textbf{Quantization loss $l_q$.}
Following \citet{defossez2022high, wu2023audiodec}, we apply a quantization loss $l_q$ between the output of the encoder and the quantized value from VQ. 
Formally, given the latent representations $\rmZ = [\rvz_1, \rvz_2, \cdots, \rvz_t]$ and the codebooks $\rmE = [\rvz_1, \rvz_2, \cdots, \rvz_K]$, we minimize
\begin{equation}
    \label{eq:quant_loss}
    l_q = \frac{1}{T} \sum_{t=1}^T \frac{1}{H} \sum_{k=1}^K  \mathbb{I}_{k}(\rvz_t)  \|\rvz_t - \rve_k\|_2^2
\end{equation}
where $\mathbb{I}_{k}(\rvz_t) \in \{0, 1\} $ is binary indicator variables indicating which of the $K$ clusters the data point $\rvz_t$ is assigned to. $\mathbb{I}_{k}(\rvz_t)= 1$ if $\rvz_t$ is assigned to cluster $k$, and $\mathbb{I}_{k}(\rvz_t)= 0$ otherwise. When RVQ is used, the quantization loss in \Cref{eq:quant_loss} is generalized as:
\begin{equation}
    l_q = \sum_{i=1}^{M}\frac{1}{T} \sum_{t=1}^T \frac{1}{H} \sum_{k=1}^K  \mathbb{I}_{k}^i(\rvz_{t}^i)  \|\rvz_{t}^i - \rve_{k}^i\|_2^2
    \label{eq:rvq_loss}
\end{equation}
where $i \in [1, M]$, denotes the $i^{th}$ quantizer of RVQ. And the superscript $i$ in $\mathbb{I}_{k}^i, \rvz_{t}^i, \rve_{k}^i$ denotes the indicator, input representation and codebook for $i^{th}$ quantizer respectively.
$l_q$ is only used for updating the encoder parameters. It makes the latent representation $\rmZ$ of the encoder suitable for the clustering of quantizer. The quantizer is updated by EMA described in \Cref{sec:kmeans_vq}.
Overall, {\rpc} is trained by a combination of the two losses,
\begin{equation}
    l = \lambda_r \cdot l_r + \lambda_q \cdot l_q
\end{equation}


\subsection{Optimization of Vector Quantizer}
\label{sec:kmeans_vq}

Both k-means \citep{lloyd1982least} and VQ \citep{gray1984vector} are algorithms to discretize a high-dimensional vector into a discrete label.
They optimize a common objective function: aiming to find the best clusters measured by $\ell_2$ in \Cref{eq:quant_loss} \citep{bishop2006pattern}.
However, they adopt different kinds of optimization algorithms. 

K-means adopts an EM algorithm \citep{bishop2006pattern} to search for the best clusters. Despite its success in clustering raw input, its sharp changes hinder the back-propagation of gradient through the quantization module, which potentially results in training process instability. 
On the contrary, the optimization algorithms adopted in VQ, including Straight-through Gradient Method, Exponential Moving Average (EMA) and Gumble-softmax, gradually change the quantization. This ensures a stable update of the encoder so that it can be trained end-to-end with other components of the model \citep{van2017vqvae}. 
We follow the optimization in \citet{soundstream} and use the EMA algorithm. 
Formally speaking, let $\{\rvz_1, \ldots, \rvz_b\}$ be the minibatch input, where $b$ is the batch size, then the codebook entries $\rve_k$ are updated by EMA with factor $0 \leq \gamma \leq 1$, where $\tilde{n}_k$ and $\tilde{e}_k$ represents the moving average of the number and codebook of the $k$-th cluster. Denoting $\mathbb{I}_k(\rvz_j)$ as the indicator that $j$-th feature belongs to $k$-th cluster, we have

\begin{equation}
    \begin{aligned}
    \tilde{n}_k &= \gamma \tilde{n}_k  + (1-\gamma) \sum_{j=1}^b \mathbb{I}_k(\rvz_j), \\
    \tilde{\rve}_k &= \gamma\tilde{\rve}_k +  (1 - \gamma) \sum_{j=1}^b \mathbb{I}_k(\rvz_j) \rvz_j
    \end{aligned}
\end{equation}


\subsection{Downstream Tasks}

To measure the performance of semantic tokens, we evaluate these tokens on downstream tasks of decoder-only ASR and unit-to-speech generation. These two sets of experiments simulate the audio input and audio output of a language model respectively. And we measure the amount of semantic information captured by the tokens by WER.

\noindent\textbf{Speech Resynthesis and Voice Conversion.}
For each set of tokens, a unit-based HiFi-GAN vocoder is built to resynthesize speech. 
We follow \citet{directs2st,polyak2021speech} for the training and inference of the vocoders. 
The vocoders are trained with a combination of the generator-discriminator loss and the mean square error (MSE) of each unit segment in logarithmic domain. 
Following the common practice to evaluate semantic tokens \citep{borsos2023audiolm, lee-etal-2022-textless}, the quality of tokens are measured by ASR-WER of the resynthesized speech with the Whisper 
large-v2 model \citep{radford2023whisper}. Additionally, the acoustic quality of the speech is evaluated by F0 error and MOS, which are presented in \Cref{tab:speaker_similarity}.

\noindent\textbf{Decoder-only ASR.} Following the emergence of LLM, and a series of works that inject audio information into a decoder-only transformer, we evaluate the quality of {\rpc} using a decoder-only transformer. Given a series of semantic audio tokens $\rvs = s_1 s_2 \cdots s_T$ and its corresponding transcript $\rvy = y_1 y_2 \cdots y_m$, we form a sequence 
\begin{equation*}
 \begin{aligned}
     \rvp
     = s_1 s_2 \cdots s_n \  \text{\texttt{<|transcribe|>}} \  y_1 y_2 \cdots y_m
 \end{aligned}
\end{equation*}
where \texttt{<|transcribe|>} is a special token indicating the start the transcription. As ASR is a sequence-to-sequence task, we find a transformer $F$ that maximizes the conditional probability 
\begin{equation*}
    F_* = \argmax_{F} p(\rvy | \rvs) = \argmax_{F} \prod_{i=1}^m p(y_i | y_{<i}, \rvs),
\end{equation*}
instead of full language modeling of $p(\rvs, \rvy)$. We also compare the full language modeling of $p(\rvs, \rvy)$ in \Cref{sec:exp_app_source}, which is much worse than $p(\rvy | \rvs)$.


\section{Experiments}
\label{sec:exp}

\subsection{Experiment Setups}

\noindent\textbf{Choices of Representation.} We select the widely-used self-supervised pretrained models, HuBERT \citep{hsu2021hubert} and data2vec \citep{baevski2022data2vec}, as the speech encoder of {\rpc}. In accordance with the common practices of selecting the layer for representations, we choose the output from the layer at about 2/3 of the total layers as the input representations of {\rpc}. In addition, we include the most-powerful open-sourced representations, Whisper \citep{radford2023whisper}, in our evaluation of {\rpc}. As we find the representations from the top layer of Whisper encoder is most suitable for ASR in SUPERB \citep{yang2021superb}, we use them as the input representations for {\rpc}.
Moreover, we extend our evaluation beyond single-layer representations, exploring whether the linear combinations of multiple layers are more suitable for the downstream tasks. We use SUPERB toolkit to find the best linear combination of representations for SUPERB ASR, and use it as the input representation for {\rpc}.

\noindent\textbf{Baselines.} We compare {\rpc} with several baselines, including \textbf{k-means} \citep{hsu2021hubert}, \textbf{VQ} and \textbf{EnCodec} \citep{defossez2022high}. \textbf{K-means} has been a predominant method in prior literature for semantic token extraction \citep{hsu2021hubert, lee-etal-2022-textless, borsos2023audiolm}. \textbf{VQ} directly takes the input speech representation $\rmX$ to the vector quantizer without either encoder or decoder. K-means and VQ are similar methods, sharing the same model and objective function, except that VQ uses the same  optimization method as {\rpc}. EnCodec is an open-source model similar to SoundStream \citep{soundstream}, which compresses the information directly from raw audio. We limit the bitrates of EnCodec for fair comparison among different methods.

In addition to the baselines, we also include a upper bound for decoder-only ASR, speech resynthesis and voice conversion respectively. For ASR, we replace the tokens with the original representations as the input of the decoder, preserving the complete information for ASR. Concerning speech resynthesis and voice conversion, ASR-WER of the original audio is reported as a benchmark.

\noindent\textbf{Training Semantic Tokenizers.} The detailed architecture and hyperparameters of training {\rpc} is available at \Cref{sec:hyper_app}. We employ a fixed cluster count of $K=1024$ for all semantic tokenizers. And we use the train-clean-100 subset of the LibriSpeech corpus \citep{panayotov2015librispeech} to train our all semantic tokenizers. It ensures a fair comparison between k-means, VQ and {\rpc} (previous implementation of k-means cannot use large amount of audio data due to memory constraint). For multilingual experiments, we further incorporate 100h subsets from MLS \citep{pratap2020mls} for French and Spanish.

\begin{table*}[!t]
    \centering
    \label{tab:main_results}
    \caption{Main results on decoder-only ASR tasks. The WER scores are evaluated on the test-clean set of LibriSpeech. K-means, VQ, and {\rpc} are trained on the train-clean-100 subset. Then we use these speech tokenizers to generate tokens for the entire 960h LibriSpeech. All the ASR models are Base transformer decoder-only models and are trained on 960h of representations or tokens.}
    \setlength\tabcolsep{4.0pt}

    \resizebox{\linewidth}{!}
    {
        \begin{tabular}{@{\extracolsep{1.5pt}}l*{12}c}
        \toprule
        \multirow{4}{*}{\diagbox[trim=l,height=4\line]{Method}{Representation}} & \multicolumn{6}{c}{\textbf{Multiple Layers (Linear Combination)}} & \multicolumn{6}{c}{\textbf{Single Layer}}\\
        \cline{2-7} \cline{8-13} & 
        \multicolumn{2}{c}{HuBERT} & \multicolumn{2}{c}{data2vec} & \multicolumn{2}{c}{Whisper} &
        \multicolumn{2}{c}{HuBERT} & \multicolumn{2}{c}{data2vec} & \multicolumn{2}{c}{Whisper} \\
        \cline{2-3} \cline{4-5} \cline{6-7} \cline{8-9} \cline{10-11} \cline{12-13} &
        base & large & base & large & medium & large & 
        base & large & base & large & medium & large \\
        & - & - & - & - & - & - & 
        9\textsuperscript{th} & 18\textsuperscript{th} & 6\textsuperscript{th} & 18\textsuperscript{th} & 24\textsuperscript{th} & 32\textsuperscript{nd}\\
        \midrule
        \textit{Representation} & \textit{3.62} & \textit{2.91} & \textit{3.06} & \textit{2.18} & \textit{4.54} & \textit{6.16} 
                & \textit{4.02} & \textit{2.81} & \textit{3.77} & \textit{2.18} & \textit{3.94} & \textit{3.96} \\
        \midrule
        EnCodec (1RVQ 0.75kbps) & \multicolumn{12}{c}{35.44} \\
        EnCodec (2RVQ 1.5kbps) & \multicolumn{12}{c}{16.53} \\
        k-means (0.5kbps) & 10.83 & 6.14 & 6.57 & 7.23 & 100+ & 100+ 
                & 6.36 & 5.00 & 5.97 & 4.55 & 9.52 & 9.97 \\
        VQ  (0.5kbps)      & 10.20 & 5.17 & 6.14 & 8.53 & 100+ & 100+
                & 6.27 & 5.19 & 6.20 & 4.68 & 24.35 & 44.43 \\
        \midrule
        {\rpc}  (0.5kbps) & \textbf{9.93} & \textbf{4.11} & \textbf{4.87} & \textbf{5.39} & \textbf{12.89} & \textbf{13.12}
                 & \textbf{5.73} & \textbf{4.02} & \textbf{5.15} & \textbf{2.87} & \textbf{5.04} & \textbf{5.01} \\
        \bottomrule
        \end{tabular}
    }
    \label{tab:asr}
\end{table*}

\begin{table*}[t]
    \centering
    \caption{WER of ASR Modelling when scaling {\rpc}, using RVQ, varying the number of clusters and applying to different languages. }

    \begin{subtable}{0.49\linewidth}
        \centering
        \setlength\tabcolsep{8.5pt}

        \caption{WER of scaled {\rpc}.}
        \begin{tabular}{l c c}
            \hline 
            & HuBERT   & data2vec   \\
            & large 18\textsuperscript{th} & large 18\textsuperscript{th} \\
            \hline
            {\rpc} (100h) & 4.03 & 2.87 \\
            {\rpc} (960h) & \bfseries 3.72 & \bfseries 2.65 \\
            \hline
        \end{tabular}
        \label{tab:large_repcodec}
    \end{subtable}
    \begin{subtable}{0.49\linewidth}
        \centering
        \setlength\tabcolsep{8.0pt}

        \caption{WER of {\rpc} using RVQ.}
        \begin{tabular}{l c c}
            \hline
            & HuBERT   & data2vec   \\
            & large 18\textsuperscript{th} & large 18\textsuperscript{th} \\
            \hline
            {\rpc} (1 VQ) & 4.03 & 2.87\\
            {\rpc} (2 RVQ) &  \bfseries 3.85 & \bfseries 2.48 \\
            \hline
        \end{tabular}
        \label{tab:asr_2rvq}
    \end{subtable}
    \vfill
    \vspace{1em}
    \begin{subtable}{0.49\linewidth}
        \caption{WER of different number of clusters. We use HuBERT large 18\textsuperscript{th} for the analysis. }
        \setlength\tabcolsep{8.5pt}

        \begin{tabular}{l c c c c}
            \hline 
            Clusters $K$ & 512 & 1024 & 2048 & 4096 \\
            \hline
            k-means & 5.50 & 5.00 & 4.71 & 4.78 \\
            VQ &    5.39 & 5.19 & 4.54 & 4.71 \\
            {\rpc}  & \bfseries 4.14 & \bfseries 4.02 & \bfseries 3.89 & \bfseries 4.03 \\
            \hline
        \end{tabular}
        \label{tab:num_cluster}
    \end{subtable}
    \hfill
    \begin{subtable}{0.49\linewidth}
        \caption{WER of speech in different languages . We use mHuBERT 11\textsuperscript{th} for the analysis. }
        \setlength\tabcolsep{9.5pt}

        \begin{tabular}{l c c c}
            \hline 
            Language & English & French & Spanish \\
            \hline
            k-means & 9.60 & 13.72 & 10.70 \\
            VQ &    10.55 & 13.83 & 10.42 \\
            {\rpc}  & \bfseries 8.52 & \bfseries 12.90 & \bfseries 9.78 \\
            \hline
        \end{tabular}
        \label{tab:multilingual}
    \end{subtable}
\end{table*}

\subsection{Decoder-Only ASR}

We convert the full 960h of LibriSpeech \citep{panayotov2015librispeech} speech into speech tokens and use them to train decoder-only ASR models. The decoder is a Base transformer with 12 layers, embedding dimension 768 and FFN dimension 3072. After training, the ASR models are then evaluated at the test-clean and dev-clean subsets of LibriSpeech. Detailed experimental setup are deferred to \Cref{sec:hyper_app}.
As shown in \Cref{tab:asr}, {\rpc} achieves much lower WER than both k-means and VQ across all representations. 
For single-layer representations, {\rpc} is particularly effective for large speech encoders such as data2vec large and Whisper. For data2vec large, {\rpc} improves WER by about 2\% in absolute value, and achieves very close performance to the original representation. In case of both Whisper medium and Whisper large models, {\rpc} improves the WER by more than 4\% in absolute terms, relatively decreasing WER by nearly 50\%. 

For linear combination of representations, we observe that they are not as suitable as the single layer representations for clustering. Nevertheless, {\rpc} still achieves large improvement in WER. Particularly, {\rpc} can produce meaningful WER for Whisper representations, while both VQ and k-means cannot successfully cluster them.
Although EnCodec \citep{defossez2022high} preserves information beyond linguistic content of the speech, its performance is inferior to {\rpc} in terms of semantic information. While EnCodec uses higher bitrates than {\rpc} (1.5kps v.s 0.5kbps), {\rpc} still achieves lower WER. It shows that {\rpc} is more suitable for downstream tasks which rely on semantic information of speech.

\noindent\textbf{Scaling {\rpc}.}  In \Cref{tab:large_repcodec}, we use all 960h data of LibriSpeech to train a larger {\rpc} model (the architecture is deferred to \Cref{sec:architecture_app}). The downstream decoder-only ASR shows that the model achieves even lower WER, validating the scaling ability of our method. In \Cref{sec:scaling_compare}, we also compared the scaling ability of {\rpc} with k-means and report the WER on test-other subset of LibriSpeech.

\noindent\textbf{Residual Vector Quantization.} In \Cref{tab:asr_2rvq}, we present the performance of {\rpc} with a 2-layer RVQ. It has higher bitrates and preserve more information of the speech representation. Therefore, {\rpc} has further improvements in the downstream decoder-only ASR task.

\noindent\textbf{Multilingual.}  We conduct experiments with representations from mHuBERT \citep{lee-etal-2022-textless} and train a unified speech tokenizer for three languages (English, French and Spanish). Then we train a unified decoder for ASR of all three languages. The WER is evaluated on the test-clean of LibriSpeech and the test sets of MLS. As shown in \Cref{tab:multilingual}, {\rpc} outperforms k-means and VQ across all three languages, demonstrating that {\rpc} can be applied to multiple languages.

\subsection{Speech Resynthesis and Voice Conversion}

We conduct the unit-to-speech resynthesis and voice conversion on two datasets: LJSpeech \citep{ljspeech2017} and VCTK \citep{vctk2017} .
LJSpeech is a single speaker English TTS corpus comprising 13,100 speech utterances, equivalent to approximately 24 hours of audio. 
VCTK is a multi-speaker English TTS corpus uttered by 109 speakers.
It comprises around 43,800 speech utterances, equivalent to approximately 44 hours of audio.
We follow the data partition in \citet{polyak2021speech}, and split the data into training sets, validation sets and test sets. 
All the audios are downsampled to 16kHz and trained with a fixed 50k training steps.
For the experiments of LJSpeech, we use the architecture and \href{https://github.com/facebookresearch/speech-resynthesis/tree/main/examples/speech_to_speech_translation}{toolkit} provided in \citet{directs2st} for the training and inference of the vocoders. 
For VCTK, we follow the model architecture and \href{https://github.com/facebookresearch/speech-resynthesis/tree/main}{toolkit} provided in \citet{polyak2021speech}. In this setup, the synthesized speech's speaker characteristics are conditioned on speaker embeddings.
In \Cref{tab:tts_results}, "Speech Resynthesis" indicates that the speaker embedding utilized for resynthesizing the speech corresponds to the ground truth speaker, while the speaker embeddings of "Voice Conversion" were randomly selected from the seen speakers.

\begin{table*}[!t]
    \centering
    \caption{ASR-WER of the speech of test set of LJSpeech and VCTK. The task is speech resynthesis and voice conversion. The ASR-WER is computed with Whisper large-v2.}
    \setlength\tabcolsep{4.0pt}

    \resizebox{\linewidth}{!}
    {
        \begin{tabular}{@{\extracolsep{1.5pt}}l*{4}c|*{3}c|*{3}c}
        \toprule
        \multirow{4}{*}{\diagbox[trim=l,height=4\line]{Method}{Representation}} & \multicolumn{4}{c|}{\textbf{LJSpeech Speech Resysthesis}} & \multicolumn{3}{c|}{\textbf{VCTK Speech Resysthesis}} & \multicolumn{3}{c}{\textbf{VCTK Voice Conversion}}\\
        \cline{2-5} \cline{6-8} \cline{9-11} & 
        HuBERT & data2vec & Whisper & Whisper & HuBERT & data2vec & Whisper & HuBERT & data2vec & Whisper\\
        & Large  &  Large & Medium & Large & Large  &  Large   & Large  & Large   &  Large  & Large \\
        & 18\textsuperscript{th} &  18\textsuperscript{th}  &  24\textsuperscript{th} & 32\textsuperscript{nd} &  18\textsuperscript{th}  &   18\textsuperscript{th}  & 32\textsuperscript{nd} &  18\textsuperscript{th}  &   18\textsuperscript{th}  &  32\textsuperscript{nd} \\
        \midrule
        \textit{Original Audio} & \multicolumn{4}{c|}{\textit{3.44}} & 
        \multicolumn{6}{c}{\textit{3.28}} \\
        \midrule
        EnCodec (1RVQ 0.75kbps) &  
        \multicolumn{4}{c|}{14.70} & \multicolumn{3}{c|}{52.67} &\multicolumn{3}{c}{-}\\
        EnCodec (2RVQ 1.5kbps) &  
        \multicolumn{4}{c|}{9.74} & 
        \multicolumn{3}{c|}{10.13} &\multicolumn{3}{c}{-}\\
        k-means (0.5kbps)& 7.61 & 9.90 & 36.02 & 100+ & 6.32 & 10.91 &  35.24 & 6.61 &  10.21 &  38.8 \\
        \midrule
        {\rpc} (0.5kbps)& 
        \textbf{4.71} & \textbf{5.25} & \textbf{5.62} & \textbf{6.18} & \textbf{4.58} & \textbf{4.88} & \textbf{6.43} & \textbf{4.41} &  \textbf{4.61} & \textbf{7.19} \\
        \bottomrule
        \end{tabular}
    }
    \label{tab:tts_results}
\end{table*}

\begin{figure*}[t]
    \centering
    \begin{minipage}{0.49\linewidth}
        \centering
        \centerline{\includegraphics[width=\linewidth]{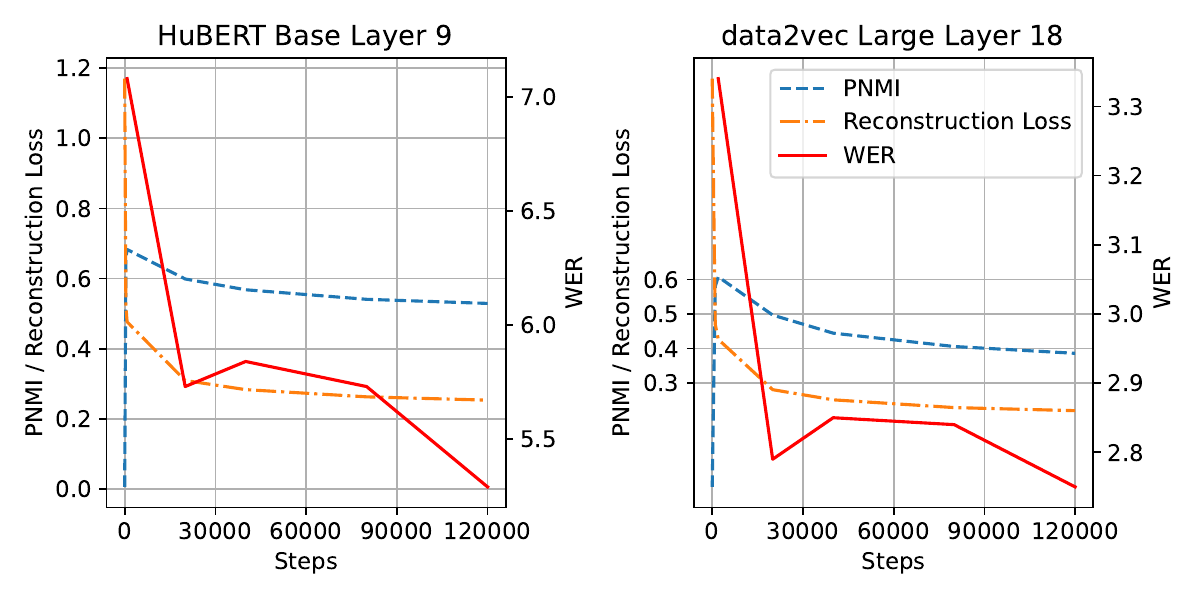}}
        \label{fig:pnmi_vs_loss_step}

    \end{minipage}
    \begin{minipage}{0.49\linewidth}
        \centering
        \centerline{\includegraphics[width=\linewidth]{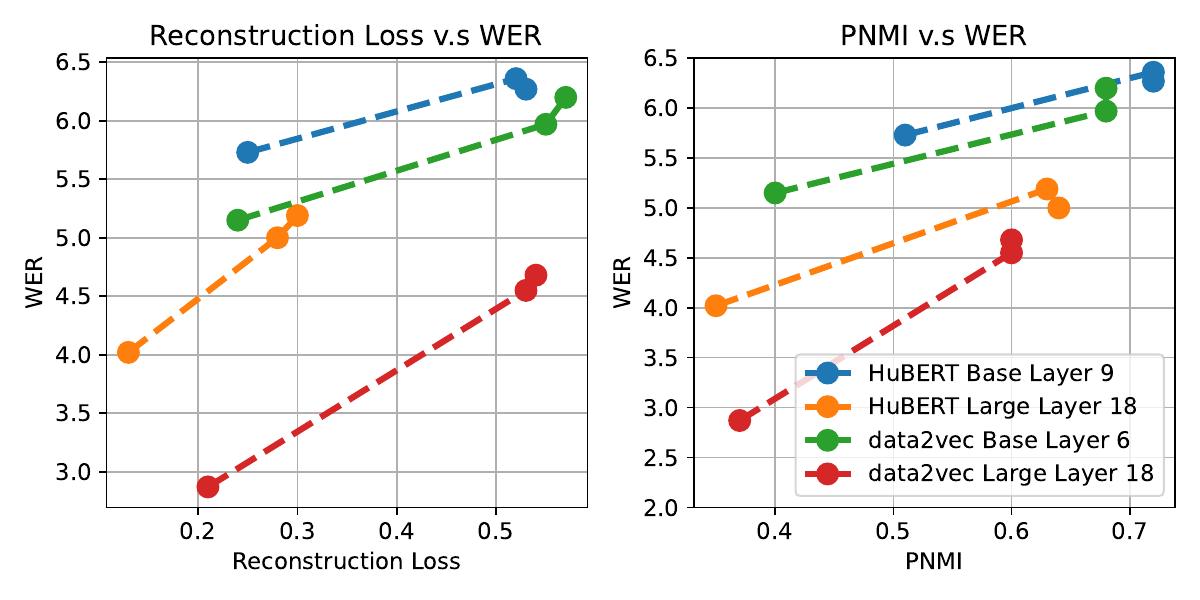}}
        \label{fig:pnmi_vs_loss_method}

    \end{minipage}

    \caption{Left: Changes of PNMI, reconstruction loss $l_r$ and WER of decoder-only ASR on test-clean of LibriSpeech as the training step of {\rpc} increases. Right: Relationship between of PNMI, reconstruction loss $l_r$, and WER of decoder-only ASR on k-means, VQ and {\rpc}.
    }
    \label{fig:pnmi_vs_loss}
\end{figure*}

\begin{table*}[!t]
    \centering
    \caption{F0 VDE, F0 FFE, MOS and Speaker Similarity of the resynthesized speech of LJSpeech.}
    \setlength\tabcolsep{4.0pt}
    \begin{tabular}{l c c c c}
        \hline 
        Method  & F0 VDE↓ & F0 FFE ↓ & Speaker Similarity ↑ & MOS \\
        \hline
        k-means & 0.199 & 0.248 & 0.771 & 2.82±0.36\\
        RepCodec & 0.174 & 0.185 & 0.764 & 3.48±0.74\\
        Original Speech  & - & - & - & 4.32±0.92 \\
        \hline
    \end{tabular}
    \label{tab:speaker_similarity}
\end{table*}

In \Cref{tab:tts_results}, we report the ASR-WER of EnCodec, {\rpc} and k-means. 
We only compare {\rpc} with k-means, which is similar to VQ and achieves lower WER in ASR. We only evaluate semantic tokens of single-layer representations, which are shown more suitable for downstream tasks in \Cref{tab:asr}. When using Whisper to transcribe the speech, we turn off the temperature scheduling and use temperature 0 to remove the randomness of the evaluation.
The WER difference between original audio and generated speech shows the quality of the semantic tokens.
{\rpc} reduces WER by more than 2\% in absolute value for all these representations in both LJSpeech and VCTK, which is much more significant than the improvement in decoder-only ASR. For Whisper representations, {\rpc} improves WERs by more than 30\% in absolute terms. This performance is in stark contrast to the deteriorations of ASR-WER in \citet{borsos2023audiolm} and \citet{lee-etal-2022-textless}. In those instances, the WER experienced a relative increase of approximately 70\%. However, for {\rpc}, this relative downgrade is significantly reduced to about 35\%.

In \Cref{tab:speaker_similarity}, we present the acoustic quality of the generated speech. Following \citet{polyak2021speech}, we use F0 Voicing Decision Error (VDE),  F0 Frame Error (FFE), Speaker Similarity and Mean
Opinion Scores (MOS) to measure the similarity and naturalness of the generates speech. The Speaker Similarity follows  \citet{wang2023wespeaker} and calculates the cosine similarity between the speaker embedding. \Cref{tab:speaker_similarity} shows that the tokens of {\rpc} does not produce downgraded quality of the speech. On the contrary, with better information preservation, {\rpc} even improves the acoustic quality of the resynthesized speech, proving the superiority of the acoustic quality of {\rpc} tokens in the speech generation tasks.


\subsection{Analysis}
\label{sec:pnmi_vs_loss}

\noindent\textbf{Phone-Normalized Mutual Information (PNMI) versus Reconstruction Loss.} \citet{hsu2021hubert} and \citet{borsos2023audiolm} propose several methods to measure the quality of semantic tokens, including PNMI and ABX error. These quantities measure the similarity between each phoneme and semantic token. Specifically, PNMI defined in \citet{hsu2021hubert} measures the correspondence of single phoneme and single semantic token. As presented in our observations, we find that such one-to-one correspondence cannot fully reflect the quality of semantic tokens.

In the left plots of \Cref{fig:pnmi_vs_loss}, we show the changes of PNMI, reconstruction loss and WER as the training step increases. The reconstruction loss is normalized by its $\ell_2$ norm. The reconstruction loss decreases as the training of {\rpc} proceeds, and so does the WER of decoder-only ASR. However, PNMI also decreases for longer training steps. 
The right of \Cref{fig:pnmi_vs_loss} shows a similar observation, where we plot the relationship of reconstruction loss v.s WER and PNMI v.s WER for k-means, VQ and {\rpc} for different speech representations. Methods with higher PNMI do not result in lower WER in downstream tasks. In contrast, downstream tasks performance is positively correlated to the reconstruction loss of the clustering.
These outcome underscores that higher PNMI does not necessarily correspond to reduced WER values. Instead, when we increasingly retain information from the speech representation (represented by decreasing $l_r$), the semantic tokens have higher quality for downstream tasks, although these tokens get dissimilar to the phonemes. 

It is worth noting that our findings do not contradict to the assertion made by \citet{hsu2021hubert}, which suggests token sets with higher PNMI lead to better performance.
In \citet{hsu2021hubert}, the discretized tokens serve as training targets for the speech encoders, while our tokens serve as representations of the speech itself for downstream tasks.
The difference in token usages leads to the diversion on the perspective of PNMI.

\begin{table}[t]
    \caption{$\text{PNMI}_n$ for k-means and {\rpc} on Hubert large 18\textsuperscript{th} feature. }
    \setlength\tabcolsep{4.5pt}

        \begin{tabular}{l c c c c}
            \hline 
            $\text{PNMI}_n$ & $\text{PNMI}_1$ & $\text{PNMI}_2$ & $\text{PNMI}_3$ & $\text{PNMI}_4$ \\
            \hline
            k-means & \bfseries 0.63 & 0.73 & 0.82 & 0.89 \\
            {\rpc}  & 0.35 & \bfseries 0.82 & \bfseries 0.99 & \bfseries 0.999 \\
            \hline
        \end{tabular}
        \label{tab:pnmi_n}
\end{table}

\noindent\textbf{Interpretability of {\rpc}.} While PNMI defined in \citet{hsu2021hubert} cannot reflect the quality of semantic tokens for downstream tasks, we believe the tokens of RepCodec still correspond to semantic tokens like phonemes. However, the correspondence is not one-to-one between individual tokens and individual phonemes. Instead, a sequence of RepCodec tokens corresponds to a sequence of phonemes. To measure the correspondence between sequences of discrete tokens and sequences of phonemes, we extend PNMI for n-gram tokens. It is calculated by the following equation
\begin{equation}
\text{PNMI}_n = \frac{I(s_j:s_{j+n}; z_{j}:z_{j+n})}{H(s_{j}:s_{j+n})},
\end{equation}
where $z_{j}:z_{j+n}$ denotes a sequence of semantic tokens with length $n$, starting at the $j$-th frame. $s_{j}:s_{j+n}$ denotes a phoneme sequence of length $n$. $I$ is mutual information and $H$ is entropy.  When $n=1$, $\text{PNMI}_1$ is the original PNMI defined in \citet{hsu2021hubert}. As shown in \Cref{tab:pnmi_n}, while $\text{PNMI}_1$ of k-means is higher than RepCodec. However, for longer sequences of tokens, RepCodec has actually higher $\text{PNMI}_n$ than k-means. This implies that for longer sequences like a word or a sentence, RepCodec provides more deterministic information for the downstream decoder to learn the corresponding text. Therefore, the performance of RepCodec is higher than k-means.

\noindent\textbf{Number of Clusters.} In \Cref{tab:num_cluster}, we study how the performance of {\rpc} for varying number of clusters $K$. With different $K$, {\rpc} all outperforms k-means and VQ. Moreover, {\rpc} is more robust against the changes of $K$ than other two baselines. Even {\rpc} with $K=512$ has lower WER than k-means with $K=4096$.

\section{Conclusion}
Interacting with LLMs through speech leads to an increased demand for speech tokenization. To this end, we propose {\rpc}, a novel speech representation codec to convert continuous speech waveforms into discretized tokens. In contrast to previous methods, {\rpc} employs a parametric network to preserve more semantic information of the speech representations. The extensive experiments demonstrate that semantic tokens extracted by {\rpc} outperform the prevalent k-means algorithm in downstream tasks of both speech understanding and generation. Moreover, the experiments also demonstrate that {\rpc} is a universal algorithm that can be applied to any speech encoders and to multiple languages.

\newpage
\section*{Limitations}
While our method obviously outperforms the baselines, there is still a performance gap between using representations and discretized tokens.  
Future works may need to use more sophisticated codec architectures (\textit{e.g.} Transformers) and objective functions (\textit{e.g.} adversarial loss) to minimize the gap. 
In addition, because of the constraint of computational resources, we cannot scale our training data to larger dataset like GigaSpeech, and we only evaluate {\rpc} on limited number of European languages like English, French and Spanish.

\bibliography{refs}

\begin{thebibliography}{45}
\expandafter\ifx\csname natexlab\endcsname\relax\def\natexlab#1{#1}\fi

\bibitem[{Baevski et~al.(2022)Baevski, Hsu, Xu, Babu, Gu, and
  Auli}]{baevski2022data2vec}
Alexei Baevski, Wei-Ning Hsu, Qiantong Xu, Arun Babu, Jiatao Gu, and Michael
  Auli. 2022.
\newblock {data2vec: A General Framework for Self-supervised Learning in
  Speech, Vision and Language}.
\newblock In \emph{Proceedings of the 39th International Conference on Machine
  Learning}, volume 162 of \emph{Proceedings of Machine Learning Research},
  pages 1298--1312. PMLR.

\bibitem[{Baevski et~al.(2020)Baevski, Zhou, Mohamed, and
  Auli}]{baevski2020wav2vec2}
Alexei Baevski, Yuhao Zhou, Abdelrahman Mohamed, and Michael Auli. 2020.
\newblock {wav2vec 2.0: A Framework for Self-Supervised Learning of Speech
  Representations}.
\newblock In \emph{Advances in Neural Information Processing Systems},
  volume~33, pages 12449--12460. Curran Associates, Inc.

\bibitem[{Bishop and Nasrabadi(2006)}]{bishop2006pattern}
Christopher~M Bishop and Nasser~M Nasrabadi. 2006.
\newblock \emph{Pattern recognition and machine learning}, volume~4.
\newblock Springer.

\bibitem[{Borsos et~al.(2023)Borsos, Marinier, Vincent, Kharitonov, Pietquin,
  Sharifi, Roblek, Teboul, Grangier, Tagliasacchi, and
  Zeghidour}]{borsos2023audiolm}
Zalán Borsos, Raphaël Marinier, Damien Vincent, Eugene Kharitonov, Olivier
  Pietquin, Matt Sharifi, Dominik Roblek, Olivier Teboul, David Grangier, Marco
  Tagliasacchi, and Neil Zeghidour. 2023.
\newblock \href {https://doi.org/10.1109/TASLP.2023.3288409} {{AudioLM: A
  Language Modeling Approach to Audio Generation}}.
\newblock \emph{IEEE/ACM Transactions on Audio, Speech, and Language
  Processing}, 31:2523--2533.

\bibitem[{Brown et~al.(2020)Brown, Mann, Ryder, Subbiah, Kaplan, Dhariwal,
  Neelakantan, Shyam, Sastry, Askell, Agarwal, Herbert-Voss, Krueger, Henighan,
  Child, Ramesh, Ziegler, Wu, Winter, Hesse, Chen, Sigler, Litwin, Gray, Chess,
  Clark, Berner, McCandlish, Radford, Sutskever, and
  Amodei}]{brown2020languagegpt3}
Tom Brown, Benjamin Mann, Nick Ryder, Melanie Subbiah, Jared~D Kaplan, Prafulla
  Dhariwal, Arvind Neelakantan, Pranav Shyam, Girish Sastry, Amanda Askell,
  Sandhini Agarwal, Ariel Herbert-Voss, Gretchen Krueger, Tom Henighan, Rewon
  Child, Aditya Ramesh, Daniel Ziegler, Jeffrey Wu, Clemens Winter, Chris
  Hesse, Mark Chen, Eric Sigler, Mateusz Litwin, Scott Gray, Benjamin Chess,
  Jack Clark, Christopher Berner, Sam McCandlish, Alec Radford, Ilya Sutskever,
  and Dario Amodei. 2020.
\newblock {Language Models are Few-Shot Learners}.
\newblock In \emph{Advances in Neural Information Processing Systems},
  volume~33, pages 1877--1901. Curran Associates, Inc.

\bibitem[{Chowdhery et~al.(2022)Chowdhery, Narang, Devlin, Bosma, Mishra,
  Roberts, Barham, Chung, Sutton, Gehrmann et~al.}]{chowdhery2022palm}
Aakanksha Chowdhery, Sharan Narang, Jacob Devlin, Maarten Bosma, Gaurav Mishra,
  Adam Roberts, Paul Barham, Hyung~Won Chung, Charles Sutton, Sebastian
  Gehrmann, et~al. 2022.
\newblock {Palm: Scaling language modeling with pathways}.
\newblock \emph{arXiv preprint arXiv:2204.02311}.

\bibitem[{Chung and Glass(2020)}]{chung2020generative}
Yu-An Chung and James Glass. 2020.
\newblock {Generative pre-training for speech with autoregressive predictive
  coding}.
\newblock In \emph{ICASSP 2020-2020 IEEE International Conference on Acoustics,
  Speech and Signal Processing (ICASSP)}, pages 3497--3501. IEEE.

\bibitem[{Chung et~al.(2021)Chung, Zhang, Han, Chiu, Qin, Pang, and
  Wu}]{chung2021w2v}
Yu-An Chung, Yu~Zhang, Wei Han, Chung-Cheng Chiu, James Qin, Ruoming Pang, and
  Yonghui Wu. 2021.
\newblock \href {https://doi.org/10.1109/ASRU51503.2021.9688253} {{w2v-BERT:
  Combining Contrastive Learning and Masked Language Modeling for
  Self-Supervised Speech Pre-Training}}.
\newblock In \emph{2021 IEEE Automatic Speech Recognition and Understanding
  Workshop (ASRU)}, pages 244--250.

\bibitem[{Conneau et~al.(2021)Conneau, Baevski, Collobert, Mohamed, and
  Auli}]{conneau2020unsupervisedXLSR}
Alexis Conneau, Alexei Baevski, Ronan Collobert, Abdelrahman Mohamed, and
  Michael Auli. 2021.
\newblock \href {https://doi.org/10.21437/Interspeech.2021-329} {{Unsupervised
  Cross-Lingual Representation Learning for Speech Recognition}}.
\newblock In \emph{Proc. Interspeech 2021}, pages 2426--2430.

\bibitem[{D{\'e}fossez et~al.(2022)D{\'e}fossez, Copet, Synnaeve, and
  Adi}]{defossez2022high}
Alexandre D{\'e}fossez, Jade Copet, Gabriel Synnaeve, and Yossi Adi. 2022.
\newblock {High Fidelity Neural Audio Compression}.
\newblock \emph{arXiv preprint arXiv:2210.13438}.

\bibitem[{Dong et~al.(2023)Dong, Huang, Xu, Zhao, Wang, Cheng, Ko, Tian, Li,
  Yue et~al.}]{dong2023polyvoice}
Qianqian Dong, Zhiying Huang, Chen Xu, Yunlong Zhao, Kexin Wang, Xuxin Cheng,
  Tom Ko, Qiao Tian, Tang Li, Fengpeng Yue, et~al. 2023.
\newblock {PolyVoice: Language Models for Speech to Speech Translation}.
\newblock \emph{arXiv preprint arXiv:2306.02982}.

\bibitem[{Gray(1984)}]{gray1984vector}
R.~Gray. 1984.
\newblock \href {https://doi.org/10.1109/MASSP.1984.1162229} {{Vector
  Quantization}}.
\newblock \emph{IEEE ASSP Magazine}, 1(2):4--29.

\bibitem[{Gulati et~al.(2020)Gulati, Qin, Chiu, Parmar, Zhang, Yu, Han, Wang,
  Zhang, Wu, and Pang}]{gulati2020conformer}
Anmol Gulati, James Qin, Chung-Cheng Chiu, Niki Parmar, Yu~Zhang, Jiahui Yu,
  Wei Han, Shibo Wang, Zhengdong Zhang, Yonghui Wu, and Ruoming Pang. 2020.
\newblock \href {https://doi.org/10.21437/Interspeech.2020-3015} {{Conformer:
  Convolution-augmented Transformer for Speech Recognition}}.
\newblock In \emph{Proc. Interspeech 2020}, pages 5036--5040.

\bibitem[{Gârbacea et~al.(2019)Gârbacea, den Oord, Li, Lim, Luebs, Vinyals,
  and Walters}]{garbacea2019low}
Cristina Gârbacea, Aäron~van den Oord, Yazhe Li, Felicia S~C Lim, Alejandro
  Luebs, Oriol Vinyals, and Thomas~C Walters. 2019.
\newblock \href {https://doi.org/10.1109/ICASSP.2019.8683277} {{Low Bit-rate
  Speech Coding with VQ-VAE and a WaveNet Decoder}}.
\newblock In \emph{ICASSP 2019 - 2019 IEEE International Conference on
  Acoustics, Speech and Signal Processing (ICASSP)}, pages 735--739.

\bibitem[{He et~al.(2016)He, Zhang, Ren, and Sun}]{he2016deep}
Kaiming He, Xiangyu Zhang, Shaoqing Ren, and Jian Sun. 2016.
\newblock {Deep Residual Learning for Image Recognition}.
\newblock In \emph{2016 IEEE Conference on Computer Vision and Pattern
  Recognition (CVPR)}, pages 770--778.

\bibitem[{Hsu et~al.(2021)Hsu, Bolte, Tsai, Lakhotia, Salakhutdinov, and
  Mohamed}]{hsu2021hubert}
Wei-Ning Hsu, Benjamin Bolte, Yao-Hung~Hubert Tsai, Kushal Lakhotia, Ruslan
  Salakhutdinov, and Abdelrahman Mohamed. 2021.
\newblock \href {https://doi.org/10.1109/TASLP.2021.3122291} {{HuBERT:
  Self-Supervised Speech Representation Learning by Masked Prediction of Hidden
  Units}}.
\newblock \emph{IEEE/ACM Transactions on Audio, Speech, and Language
  Processing}, 29:3451--3460.

\bibitem[{Ito and Johnson(2017)}]{ljspeech2017}
K.~Ito and L.~Johnson. 2017.
\newblock \href {https://keithito. com/LJ-Speech-Dataset/} {{The LJ Speech
  Dataset}}.

\bibitem[{Kankanahalli(2018)}]{kankanahalli2018end}
Srihari Kankanahalli. 2018.
\newblock \href {https://doi.org/10.1109/ICASSP.2018.8461487} {{End-To-End
  Optimized Speech Coding with Deep Neural Networks}}.
\newblock In \emph{2018 IEEE International Conference on Acoustics, Speech and
  Signal Processing (ICASSP)}, pages 2521--2525.

\bibitem[{Kingma and Ba(2014)}]{kingma2014adam}
Diederik~P Kingma and Jimmy Ba. 2014.
\newblock Adam: A method for stochastic optimization.
\newblock \emph{arXiv preprint arXiv:1412.6980}.

\bibitem[{Kingma and Welling(2014)}]{kingma2014vae}
Diederik~P. Kingma and Max Welling. 2014.
\newblock \href {http://arxiv.org/abs/http://arxiv.org/abs/1312.6114v10}
  {{Auto-Encoding Variational Bayes}}.
\newblock In \emph{2nd International Conference on Learning Representations,
  {ICLR} 2014, Banff, AB, Canada, April 14-16, 2014, Conference Track
  Proceedings}.

\bibitem[{Kudo and Richardson(2018)}]{kudo2018sentencepiece}
Taku Kudo and John Richardson. 2018.
\newblock {S}entence{P}iece: A simple and language independent subword
  tokenizer and detokenizer for neural text processing.
\newblock In \emph{Proceedings of the 2018 Conference on Empirical Methods in
  Natural Language Processing: System Demonstrations}.

\bibitem[{Lee et~al.(2022{\natexlab{a}})Lee, Chen, Wang, Gu, Popuri, Ma,
  Polyak, Adi, He, Tang, Pino, and Hsu}]{directs2st}
Ann Lee, Peng-Jen Chen, Changhan Wang, Jiatao Gu, Sravya Popuri, Xutai Ma, Adam
  Polyak, Yossi Adi, Qing He, Yun Tang, Juan Pino, and Wei-Ning Hsu.
  2022{\natexlab{a}}.
\newblock {Direct Speech-to-Speech Translation With Discrete Units}.
\newblock In \emph{Proceedings of the 60th Annual Meeting of the Association
  for Computational Linguistics (Volume 1: Long Papers)}, pages 3327--3339.

\bibitem[{Lee et~al.(2022{\natexlab{b}})Lee, Gong, Duquenne, Schwenk, Chen,
  Wang, Popuri, Adi, Pino, Gu, and Hsu}]{lee-etal-2022-textless}
Ann Lee, Hongyu Gong, Paul-Ambroise Duquenne, Holger Schwenk, Peng-Jen Chen,
  Changhan Wang, Sravya Popuri, Yossi Adi, Juan Pino, Jiatao Gu, and Wei-Ning
  Hsu. 2022{\natexlab{b}}.
\newblock Textless speech-to-speech translation on real data.
\newblock In \emph{Proceedings of the 2022 Conference of the North American
  Chapter of the Association for Computational Linguistics: Human Language
  Technologies}.

\bibitem[{Lloyd(1982)}]{lloyd1982least}
S.~Lloyd. 1982.
\newblock \href {https://doi.org/10.1109/TIT.1982.1056489} {{Least Squares
  Quantization in PCM}}.
\newblock \emph{IEEE Transactions on Information Theory}, 28(2):129--137.

\bibitem[{Meng et~al.(2023)Meng, Ao, Ko, Wang, and Li}]{meng2023cobert}
Chutong Meng, Junyi Ao, Tom Ko, Mingxuan Wang, and Haizhou Li. 2023.
\newblock \href {https://doi.org/10.21437/Interspeech.2023-1390} {{CoBERT:
  Self-Supervised Speech Representation Learning Through Code Representation
  Learning}}.
\newblock In \emph{Proc. INTERSPEECH 2023}, pages 2978--2982.

\bibitem[{Nguyen et~al.(2022)Nguyen, Sagot, and Dupoux}]{nguyen2022discrete}
Tu~Anh Nguyen, Benoit Sagot, and Emmanuel Dupoux. 2022.
\newblock Are discrete units necessary for spoken language modeling?
\newblock \emph{IEEE Journal of Selected Topics in Signal Processing},
  16(6):1415--1423.

\bibitem[{OpenAI(2023)}]{openai2023gpt4}
OpenAI. 2023.
\newblock \href {http://arxiv.org/abs/2303.08774} {{GPT-4 Technical Report}}.

\bibitem[{Ott et~al.(2019)Ott, Edunov, Baevski, Fan, Gross, Ng, Grangier, and
  Auli}]{ott2019fairseq}
Myle Ott, Sergey Edunov, Alexei Baevski, Angela Fan, Sam Gross, Nathan Ng,
  David Grangier, and Michael Auli. 2019.
\newblock fairseq: A fast, extensible toolkit for sequence modeling.
\newblock In \emph{Proceedings of NAACL-HLT 2019: Demonstrations}.

\bibitem[{Panayotov et~al.(2015)Panayotov, Chen, Povey, and
  Khudanpur}]{panayotov2015librispeech}
Vassil Panayotov, Guoguo Chen, Daniel Povey, and Sanjeev Khudanpur. 2015.
\newblock \href {https://doi.org/10.1109/ICASSP.2015.7178964} {{Librispeech: An
  ASR corpus based on public domain audio books}}.
\newblock In \emph{2015 IEEE International Conference on Acoustics, Speech and
  Signal Processing (ICASSP)}, pages 5206--5210.

\bibitem[{Polyak et~al.(2021)Polyak, Adi, Copet, Kharitonov, Lakhotia, Hsu,
  Mohamed, and Dupoux}]{polyak2021speech}
Adam Polyak, Yossi Adi, Jade Copet, Eugene Kharitonov, Kushal Lakhotia,
  Wei-Ning Hsu, Abdelrahman Mohamed, and Emmanuel Dupoux. 2021.
\newblock {Speech Resynthesis from Discrete Disentangled Self-Supervised
  Representations}.
\newblock In \emph{Proc. of INTERSPEECH}.

\bibitem[{Pratap et~al.(2020)Pratap, Xu, Sriram, Synnaeve, and
  Collobert}]{pratap2020mls}
Vineel Pratap, Qiantong Xu, Anuroop Sriram, Gabriel Synnaeve, and Ronan
  Collobert. 2020.
\newblock {MLS: A Large-Scale Multilingual Dataset for Speech Research}.
\newblock In \emph{Proc. Interspeech 2020}, pages 2757--2761.

\bibitem[{Radford et~al.(2023)Radford, Kim, Xu, Brockman, Mcleavey, and
  Sutskever}]{radford2023whisper}
Alec Radford, Jong~Wook Kim, Tao Xu, Greg Brockman, Christine Mcleavey, and
  Ilya Sutskever. 2023.
\newblock {Robust Speech Recognition via Large-Scale Weak Supervision}.
\newblock In \emph{Proceedings of the 40th International Conference on Machine
  Learning}, volume 202, pages 28492--28518. PMLR.

\bibitem[{Radford et~al.(2019)Radford, Wu, Child, Luan, Amodei, Sutskever
  et~al.}]{radford2019languagegpt2}
Alec Radford, Jeffrey Wu, Rewon Child, David Luan, Dario Amodei, Ilya
  Sutskever, et~al. 2019.
\newblock Language models are unsupervised multitask learners.
\newblock \emph{OpenAI blog}, 1(8):9.

\bibitem[{Rubenstein et~al.(2023)Rubenstein, Asawaroengchai, Nguyen, Bapna,
  Borsos, de~Chaumont~Quitry, Chen, Badawy, Han, Kharitonov, Muckenhirn,
  Padfield, Qin, Rozenberg, Sainath, Schalkwyk, Sharifi, Ramanovich,
  Tagliasacchi, Tudor, Velimirović, Vincent, Yu, Wang, Zayats, Zeghidour,
  Zhang, Zhang, Zilka, and Frank}]{rubenstein2023audiopalm}
Paul~K. Rubenstein, Chulayuth Asawaroengchai, Duc~Dung Nguyen, Ankur Bapna,
  Zalán Borsos, Félix de~Chaumont~Quitry, Peter Chen, Dalia~El Badawy, Wei
  Han, Eugene Kharitonov, Hannah Muckenhirn, Dirk Padfield, James Qin, Danny
  Rozenberg, Tara Sainath, Johan Schalkwyk, Matt Sharifi, Michelle~Tadmor
  Ramanovich, Marco Tagliasacchi, Alexandru Tudor, Mihajlo Velimirović, Damien
  Vincent, Jiahui Yu, Yongqiang Wang, Vicky Zayats, Neil Zeghidour, Yu~Zhang,
  Zhishuai Zhang, Lukas Zilka, and Christian Frank. 2023.
\newblock \href {http://arxiv.org/abs/2306.12925} {{AudioPaLM: A Large Language
  Model That Can Speak and Listen}}.

\bibitem[{van~den Oord et~al.(2017)van~den Oord, Vinyals, and
  kavukcuoglu}]{van2017vqvae}
Aaron van~den Oord, Oriol Vinyals, and koray kavukcuoglu. 2017.
\newblock {Neural Discrete Representation Learning}.
\newblock In \emph{Advances in Neural Information Processing Systems},
  volume~30. Curran Associates, Inc.

\bibitem[{Veaux et~al.(2017)Veaux, Yamagishi, and MacDonald}]{vctk2017}
Christophe Veaux, Junichi Yamagishi, and Kirsten MacDonald. 2017.
\newblock {CSTR VCTK Corpus: English Multi-speaker Corpus for CSTR Voice
  Cloning Toolkit}.

\bibitem[{Wang et~al.(2023{\natexlab{a}})Wang, Chen, Wu, Zhang, Zhou, Liu,
  Chen, Liu, Wang, Li et~al.}]{wang2023neural}
Chengyi Wang, Sanyuan Chen, Yu~Wu, Ziqiang Zhang, Long Zhou, Shujie Liu, Zhuo
  Chen, Yanqing Liu, Huaming Wang, Jinyu Li, et~al. 2023{\natexlab{a}}.
\newblock {Neural Codec Language Models are Zero-Shot Text to Speech
  Synthesizers}.
\newblock \emph{arXiv preprint arXiv:2301.02111}.

\bibitem[{Wang et~al.(2023{\natexlab{b}})Wang, Liang, Wang, Chen, Zhang, Xiang,
  Deng, and Qian}]{wang2023wespeaker}
Hongji Wang, Chengdong Liang, Shuai Wang, Zhengyang Chen, Binbin Zhang,
  Xu~Xiang, Yanlei Deng, and Yanmin Qian. 2023{\natexlab{b}}.
\newblock Wespeaker: A research and production oriented speaker embedding
  learning toolkit.
\newblock In \emph{IEEE International Conference on Acoustics, Speech and
  Signal Processing (ICASSP)}, pages 1--5. IEEE.

\bibitem[{Wei et~al.(2021)Wei, Bosma, Zhao, Guu, Yu, Lester, Du, Dai, and
  Le}]{wei2021finetunedflan}
Jason Wei, Maarten Bosma, Vincent Zhao, Kelvin Guu, Adams~Wei Yu, Brian Lester,
  Nan Du, Andrew~M Dai, and Quoc~V Le. 2021.
\newblock {Finetuned Language Models are Zero-Shot Learners}.
\newblock In \emph{International Conference on Learning Representations}.

\bibitem[{wen Yang et~al.(2021)wen Yang, Chi, Chuang, Lai, Lakhotia, Lin, Liu,
  Shi, Chang, Lin, Huang, Tseng, tik Lee, Liu, Huang, Dong, Li, Watanabe,
  Mohamed, and yi~Lee}]{yang2021superb}
Shu wen Yang, Po-Han Chi, Yung-Sung Chuang, Cheng-I~Jeff Lai, Kushal Lakhotia,
  Yist~Y. Lin, Andy~T. Liu, Jiatong Shi, Xuankai Chang, Guan-Ting Lin,
  Tzu-Hsien Huang, Wei-Cheng Tseng, Ko~tik Lee, Da-Rong Liu, Zili Huang, Shuyan
  Dong, Shang-Wen Li, Shinji Watanabe, Abdelrahman Mohamed, and Hung yi~Lee.
  2021.
\newblock \href {https://doi.org/10.21437/Interspeech.2021-1775} {{SUPERB:
  Speech Processing Universal PERformance Benchmark}}.
\newblock In \emph{Proc. Interspeech 2021}, pages 1194--1198.

\bibitem[{Wu et~al.(2023)Wu, Gebru, Markovi{\'c}, and Richard}]{wu2023audiodec}
Yi-Chiao Wu, Israel~D Gebru, Dejan Markovi{\'c}, and Alexander Richard. 2023.
\newblock Audiodec: An open-source streaming high-fidelity neural audio codec.
\newblock In \emph{ICASSP 2023-2023 IEEE International Conference on Acoustics,
  Speech and Signal Processing (ICASSP)}, pages 1--5. IEEE.

\bibitem[{Yang et~al.(2022)Yang, Marković, Krenn, Agrawal, and
  Richard}]{yang2022audio}
Karren Yang, Dejan Marković, Steven Krenn, Vasu Agrawal, and Alexander
  Richard. 2022.
\newblock \href {https://doi.org/10.1109/CVPR52688.2022.00805} {{Audio-Visual
  Speech Codecs: Rethinking Audio-Visual Speech Enhancement by Re-Synthesis}}.
\newblock In \emph{2022 IEEE/CVF Conference on Computer Vision and Pattern
  Recognition (CVPR)}, pages 8217--8227.

\bibitem[{Zeghidour et~al.(2022)Zeghidour, Luebs, Omran, Skoglund, and
  Tagliasacchi}]{soundstream}
Neil Zeghidour, Alejandro Luebs, Ahmed Omran, Jan Skoglund, and Marco
  Tagliasacchi. 2022.
\newblock \href {https://doi.org/10.1109/TASLP.2021.3129994} {{SoundStream: An
  End-to-End Neural Audio Codec}}.
\newblock \emph{IEEE/ACM Transactions on Audio, Speech, and Language
  Processing}, 30:495--507.

\bibitem[{Zhang et~al.(2023{\natexlab{a}})Zhang, Li, Zhang, Zhan, Wang, Zhou,
  and Qiu}]{zhang2023speechgpt}
Dong Zhang, Shimin Li, Xin Zhang, Jun Zhan, Pengyu Wang, Yaqian Zhou, and
  Xipeng Qiu. 2023{\natexlab{a}}.
\newblock Speech{GPT}: Empowering {L}arge {L}anguage {M}odels with {I}ntrinsic
  {C}ross-{M}odal {C}onversational {A}bilities.
\newblock \emph{arXiv preprint arXiv:2305.11000}.

\bibitem[{Zhang et~al.(2023{\natexlab{b}})Zhang, Zhang, Li, Zhou, and
  Qiu}]{zhang2023speechtokenizer}
Xin Zhang, Dong Zhang, Shimin Li, Yaqian Zhou, and Xipeng Qiu.
  2023{\natexlab{b}}.
\newblock Speechtokenizer: Unified speech tokenizer for speech large language
  models.
\newblock \emph{arXiv preprint arXiv:2308.16692}.

\end{thebibliography}
\bibliographystyle{acl_natbib}

\newpage

\begin{figure*}[!t]
    \centering
    \includegraphics[width=0.75\linewidth]{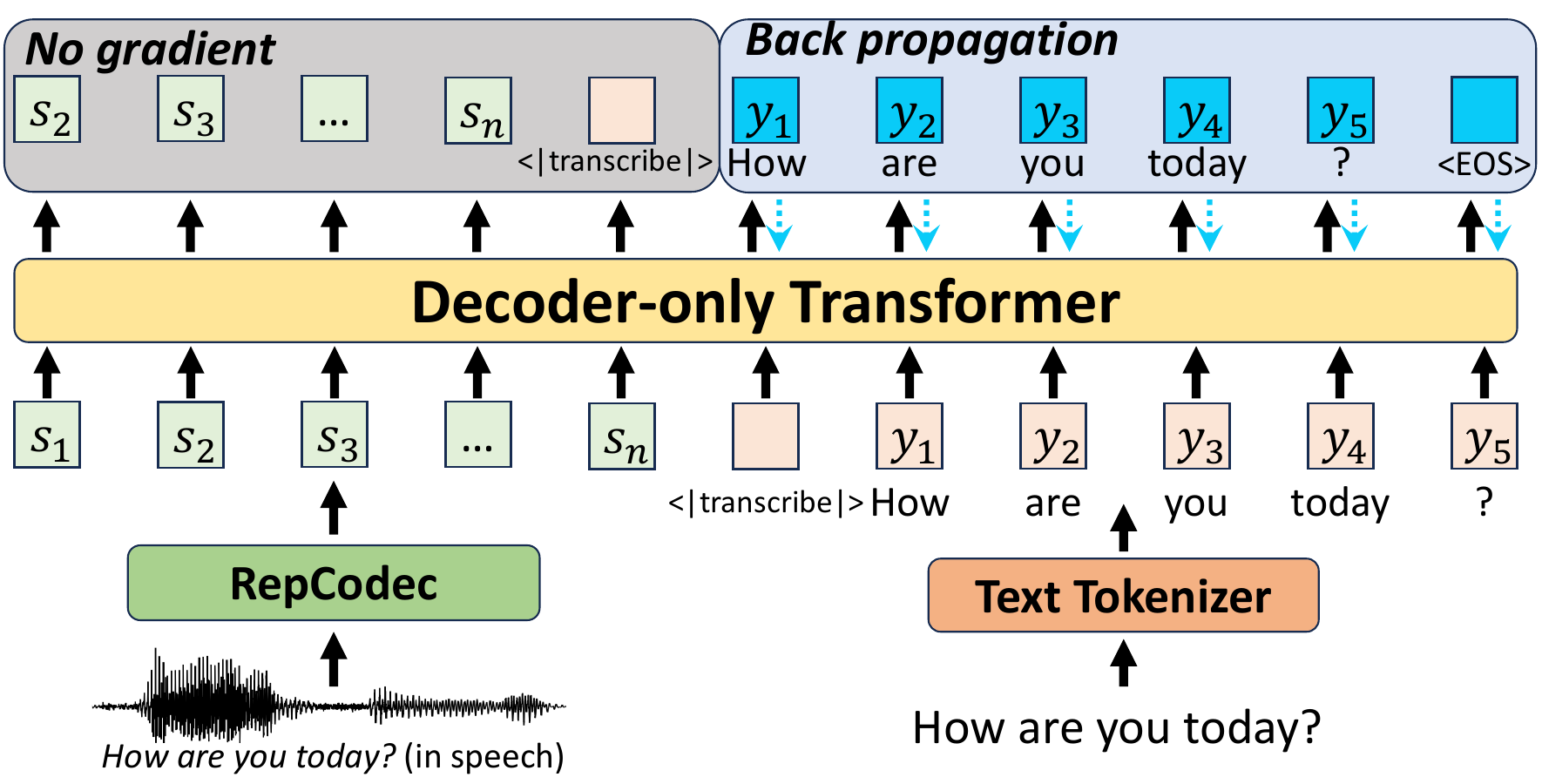}
    \caption{
    Illustration of decoder-only ASR using decoder-only Transformer architecture. Speech is tokenized by speech tokenizers and text is tokenized by SentencePiece \citep{kudo2018sentencepiece}.
    During training, we only compute gradients and apply back propagation on the text tokens (in \textcolor{cyan}{blue}).
    }
    \label{fig:language_model}
\end{figure*}

\begin{table*}[t!]
    \centering
    \caption{Architecture of {\rpc}. $H$ is the dimension of the corresponding speech representation.}
    \setlength\tabcolsep{1.0pt}

    \begin{tabular}{c c c c}
        \hline
        & Regular {\rpc} & {\rpc} in \Cref{tab:asr_2rvq} & {\rpc} in \Cref{tab:large_repcodec} \\\hline
        \multirow{3}*{Encoder} & \texttt{Conv1d(H, H, 3, 1)} & \texttt{Conv1d(H, H, 3, 1)} & \texttt{Conv1d(H, H, 3, 1)} \\
        & \texttt{Enc(H, 3, 1)} $\times 2$ & \texttt{Enc(H, 3, 1)} $\times 2$ & \texttt{Enc(H, 3, 1)} $\times 8$ \\
        & \texttt{Conv1d(H, H, 3, 1)} & \texttt{Conv1d(H, H, 3, 1)} & \texttt{Conv1d(H, H, 3, 1)} \\ \hline
        \multirow{1}*{Vector Quantizer} & \texttt{RVQ(1, 1024, H)} & \texttt{RVQ(2, 1024, H)} & \texttt{RVQ(1, 1024, H)} \\ \hline
        \multirow{3}*{Decoder} & \texttt{Conv1d(H, H, 3, 1)} & \texttt{Conv1d(H, H, 3, 1)} & \texttt{Conv1d(H, H, 3, 1)} \\
        & \texttt{Dec(H, 3, 1)} $\times 2$ & \texttt{Dec(H, 3, 1)} $\times 2$ & \texttt{Dec(H, 3, 1)} $\times 2$ \\
        & \texttt{Conv1d(H, H, 3, 1)} & \texttt{Conv1d(H, H, 3, 1)} & \texttt{Conv1d(H, H, 3, 1)}  \\ 
        \hline
    \end{tabular}
    \label{tab:architecture}
\end{table*}

\appendix

\section{Detailed Architecture of {\rpc}}
\label{sec:architecture_app}

In this section, we will introduce the architecture of {\rpc}. We first provide the components of {\rpc} and then show the specific architecture used for our method in \Cref{tab:architecture}.

\noindent\textbf{1D Convolution Layer \texttt{Conv1d(C, c, k, s)}.} We denote \texttt{Conv1d(C, c, k, s)} as the non-causal 1D convolution layer with input channel $C$, output channel $c$, kernel size $k$ and stride $s$ without dilation. 

\noindent\textbf{Residual Unit \texttt{Res(C, k, s)}. } We denote \texttt{Res(C, k, s)} as the residual unit used for both the encoder block and decoder block in \Cref{fig:repcodec_arch}. Each \texttt{Res(C, k, s)} consists of two \texttt{Conv1d(C, C, k, s)} layers, with a residual path added across them.

\noindent\textbf{Encoder Block \texttt{Enc(C, k, s)}.} With the input having a channel of $C$, the encoder block \texttt{Enc(C, k, s)} consists of two Residual Units followed by a 1D convolution layer: 

\noindent\texttt{Res(C, k, s) -> Res(C, k, s)} 
\texttt{-> Conv1d(C, C, k, s)}

\noindent\textbf{Decoder Block \texttt{Dec(C, k, s)}. } With the input having a channel of $C$, the decoder block \texttt{Dec(C, k, s)} consists of a 1D convolution layer followed by two Residual Units: 

\noindent\texttt{Conv1d(C, C, k, s) -> Res(C, k, s) -> Res(C, k, s)}

\noindent\textbf{Residual Vector Quantizer \texttt{RVQ(M, K, C)}. } \texttt{RVQ(M, K, c)} denotes a $M$-layer Residual Vector Quantizer (RVQ) with number of clusters $K$ and codebook dimension $C$. When $M=1$, RVQ is equivalent to VQ.

\begin{table*}[!t]
    \centering
    \caption{The WER on the dev-clean subset of LibriSpeech.}
    \setlength\tabcolsep{4.0pt}

    \resizebox{\linewidth}{!}
    {
        \begin{tabular}{@{\extracolsep{1.5pt}}l*{12}c}
        \toprule
        \multirow{4}{*}{\diagbox[trim=l,height=4\line]{Method}{Representation}} & \multicolumn{6}{c}{\textbf{Multiple Layers (Linear Combination)}} & \multicolumn{6}{c}{\textbf{Single Layer}}\\
        \cline{2-7} \cline{8-13} & 
        \multicolumn{2}{c}{HuBERT} & \multicolumn{2}{c}{data2vec} & \multicolumn{2}{c}{Whisper} &
        \multicolumn{2}{c}{HuBERT} & \multicolumn{2}{c}{data2vec} & \multicolumn{2}{c}{Whisper} \\
        \cline{2-3} \cline{4-5} \cline{6-7} \cline{8-9} \cline{10-11} \cline{12-13} &
        base & large & base & large & medium & large & 
        base & large & base & large & medium & large \\
        & - & - & - & - & - & - & 
        9\textsuperscript{th} & 18\textsuperscript{th} & 6\textsuperscript{th} & 18\textsuperscript{th} & 24\textsuperscript{th} & 32\textsuperscript{nd}\\
        \midrule
        \textit{Representation} & 3.27 & 2.76 & 2.79 & 2.05 & 4.16 & 5.39 
                & 3.90 & 2.65 & 3.37 & 2.13 & 3.74 & 3.58 \\
        \midrule
        EnCodec (1RVQ 0.75kbps) & \multicolumn{12}{c}{36.39} \\
        EnCodec (2RVQ 1.5kbps) & \multicolumn{12}{c}{16.78} \\
        k-means (0.5kbps) & 10.91 & 5.32 & 6.39 & 7.36 & 100+ & 100+ 
                & 6.04 & 4.99 & 6.26 & 4.59 & 9.68 & 10.15 \\
        VQ  (0.5kbps)      & 10.59 & 4.98 & 5.92 & 8.46 & 100+ & 100+
                & 6.02 & 5.18 & 6.80 & 4.84 & 34.57 & 44.29 \\
        \midrule
        {\rpc}  (0.5kbps) & \textbf{9.38} & \textbf{3.96} & \textbf{4.71} & \textbf{5.01} & \textbf{12.03} & \textbf{12.36}
                 & \textbf{5.40} & \textbf{3.76} & \textbf{5.04} & \textbf{2.12} & \textbf{4.72} & \textbf{4.75} \\
        \bottomrule
        \end{tabular}
    }
    \label{tab:asr_dev}
\end{table*}

\begin{table*}[t]
    \centering
    \caption{The WER on the dev-clean subset of LibriSpeech.}
    \setlength\tabcolsep{4.0pt}
        \begin{tabular}{@{\extracolsep{1.5pt}}l*{4}c}
        \toprule
         & \multicolumn{2}{c}{\textbf{HuBERT large 18\textsuperscript{th}}} & \multicolumn{2}{c}{\textbf{data2vec large 18\textsuperscript{th}}}\\
        \cline{2-3} \cline{4-5} &  test-clean & dev-clean & test-clean & dev-clean \\
        \midrule
        k-means $p(\rvy | \rvs)$ & 5.00 & 4.99 & 4.55 & 4.59\\
        k-means $p(\rvs, \rvy)$ & 7.17 & 7.10 & 9.45 & 8.47 \\
        \midrule
        VQ $p(\rvy | \rvs)$ & 5.19 & 5.18 & 4.68 & 4.84  \\
        VQ $p(\rvs, \rvy)$ & 8.85 & 8.65 & 8.32 & 8.05  \\
        \midrule
        {\rpc} $p(\rvy | \rvs)$ & 4.02 & 3.76 & 2.87 & 2.12  \\
        {\rpc} $p(\rvs, \rvy)$ & 6.70 & 6.47 & 6.77 & 6.30  \\
        \bottomrule
        \end{tabular}
    \label{tab:exp_app_source}
    
\end{table*}

\begin{table*}[t]
    \centering
    \caption{The WER of {\rpc} and k-means trained on full LibriSpeech 960h.}
    \setlength\tabcolsep{4.0pt}
        \begin{tabular}{@{\extracolsep{1.5pt}}l*{4}c}
        \toprule
         & \multicolumn{2}{c}{\textbf{HuBERT large 18\textsuperscript{th}}} & \multicolumn{2}{c}{\textbf{data2vec large 18\textsuperscript{th}}}\\
        \cline{2-3} \cline{4-5} &  test-clean & test-other & test-clean & test-other \\
        \midrule
        k-means & 4.70 & 10.41 & 4.80 & 7.68 \\
        {\rpc} & 3.72 & 9.24 & 2.65 & 5.84\\
        \bottomrule
        \end{tabular}
    \label{tab:scaling_compare}
    
\end{table*}

\begin{table}[t]
    \centering
    \caption{WER of dev sets of multilingual experiments.}
    \setlength\tabcolsep{7.5pt}
    \begin{tabular}{l c c c}
        \hline 
        Language & English & French & Spanish \\
        \hline
        k-means & 9.70 & 15.89 & 10.21 \\
        VQ &    10.76 & 15.65 & 10.35 \\
        {\rpc}  & \bfseries 8.78 & \bfseries 14.97 & \bfseries 9.42 \\
        \hline
    \end{tabular}
    \label{tab:multilingual_dev}
\end{table}

\begin{table}[t]
    \centering
    \caption{WER, PNMI and Reconstruction Loss of decoder only ASR modeling from 18th layer of large data2vec model.}
    \setlength\tabcolsep{4.0pt}
    \begin{tabular}{l c c c c}
        \hline 
        $\lambda_q$ & $\lambda_r$ & PNMI & Reconstruction Loss & WER \\
        \hline
        1.0 & 30.0 & 0.3689 & 0.2091 & 2.91\\
        1.0 & 45.0 & 0.3714 & 0.2100 & 2.87\\
        1.0  & 60.0 & 0.3697 & 0.2090 & 2.83 \\
        \hline
    \end{tabular}
    \label{tab:change_weight}
\end{table}

\section{Details of the Experiments}
\label{sec:hyper_app}

\subsection{Training Speech Tokenizers.}

\noindent\textbf{K-means.} We use the script from HuBERT\footnote{\url{https://github.com/facebookresearch/fairseq/tree/main/examples/hubert/simple_kmeans}} to train the k-means model and perform k-means clustering, with all hyperparemeters unchanged.

\noindent\textbf{\rpc.} We train {\rpc} for 200,000 steps. The batch size is 32 speech representations, each of which has 96 frames. We use Adam \citep{kingma2014adam} to optimize the model with a fixed learning rate $1 \times 10^{-4}$ and $\beta_1 = 0.5, \beta_2 =0.9$. We set $\lambda_r = 45, \lambda_q = 1$ and weight decay as $0$ for all the experiments. 

\noindent\textbf{VQ.} We remove the encoder and decoder from {\rpc} and train the model for 50,000 steps. Other hyperparameters are the same as {\rpc}.

\subsection{Decoder-only ASR.} As shown in \Cref{fig:language_model}, we use the Base Transformer decoder from fairseq \citep{ott2019fairseq} for decoder-only ASR. We fix the training steps to 100,000 for all the experiments. The decoder is optimized by Adam with $\beta_1=0.9, \beta_2=0.999$. The learning rate is warmed up by 5,000 steps to $1 \times 10^{-3}$ and then follow inverse square root decay to $0$. We use cross entropy loss with smooth factor $0.1$ to optimize the model, and we select the best checkpoint on the dev-clean set for further evaluation on the test sets.

We use SentencePiece as the text tokenizer, and we train a new SentencePiece model on each dataset with all its transcripts.
The vocabulary size is 5,000 for English dataset.
For multilingual dataset in \Cref{tab:multilingual}, we jointly train the SentencePiece model of all three languages with  a vocabulary size of 10,000.
We implement beam search as the decoding strategy with size 5. For convenience, we do not add additional language model or length penalty. 



\section{Additional Experiments}
\label{sec:exp_app}

\subsection{Addition Results of Decoder-Only ASR.}

\Cref{tab:asr_dev} and \Cref{tab:multilingual_dev} show the WER of ASR on the dev sets. {\rpc} still outperforms baselines by a large margin. It is worth noting the {\rpc} achieves even lower WER on the layer 18 representation of data2vec large.

\subsection{Full Language Modelling v.s Conditional Language Modelling}
\label{sec:exp_app_source}

\Cref{tab:exp_app_source} compares the WER on the test-clean and dev-clean subset of LibriSpeech when optimizing $p(\rvs, \rvy)$ or $p(\rvy | \rvs)$ for decoder-only ASR. For all three methods, full language modelling $p(\rvs, \rvy)$ results in much higher WER than the conditional language modelling $p(\rvy | \rvs)$. Therefore, we choose to optimize $p(\rvy | \rvs)$ for decoder-only ASR task.

\change{
\subsection{Ablation Study on $\lambda_q$ and $\lambda_r$}

As shown in \Cref{tab:change_weight}, we change the weights of reconstruction loss $\lambda_r$ to train the encoder of RepCodec and report its WER on the downstream task of ASR modeling with representation from the 18th layer of data2vec large model. The results show that RepCodec is robust to the changes of the weights of $(\lambda_q, \lambda_r)$.

}

\subsection{Comparison of Scaling Ability of {\rpc} and k-means}
\label{sec:scaling_compare}

In \Cref{tab:scaling_compare}, we present the results of downstream tasks using tokens generated by k-means and RepCodec that are both trained on full LibriSpeech 960h. We report WERs of the decoder-only ASR on test-clean and test-other subsets. When training on a larger amount of audio, RepCodec can achieve lower WERs than that trained on the train-clean-100 subset. However, we find that k-means is not very scalable because of its limited number of parameters. Even with larger training datasets, the WER of k-means doesn't show improvement over data2vec features, and RepCodec outperforms k-means by a larger margin than the model trained on train-clean-100 subset.

\end{document}